\documentclass[11pt,a4paper]{article}
\usepackage[utf8]{inputenc}
\usepackage[T1]{fontenc}
\usepackage{amsmath,amssymb}
\usepackage[dvipsnames]{xcolor}
\usepackage{graphicx}
\usepackage{booktabs}
\usepackage{subcaption}
\usepackage{siunitx}
\usepackage{hyperref}
\usepackage{xspace}
\usepackage{mhchem}
\usepackage{lineno}
\usepackage{float}
\usepackage{placeins}
\usepackage{authblk}

\usepackage{tikz}
\usetikzlibrary{calc}
\usetikzlibrary{arrows.meta,positioning}

\title{AIGOR: A Modular, Event-Driven Neuromorphic Architecture for Configurable SNN Inference}

\author[1]{Pierpaolo Perticaroli\thanks{Corresponding authors:
  \texttt{pierpaolo.perticaroli@roma1.infn.it},
  \texttt{alessandro.lonardo@roma1.infn.it}}}
\author[2]{Roberto Ammendola}
\author[1]{Andrea Biagioni}
\author[1]{Ottorino Frezza}
\author[1]{Francesca Lo Cicero}
\author[1]{Michele Martinelli}
\author[1]{Pier Stanislao Paolucci}
\author[1]{Elena Pastorelli}
\author[1]{Luca Pontisso}
\author[3,1]{Cristian Rossi}
\author[1]{Francesco Simula}
\author[1]{Piero Vicini}
\author[1]{Alessandro Lonardo}

\affil[1]{Istituto Nazionale di Fisica Nucleare, Sezione di Roma, Rome, Italy}
\affil[2]{Istituto Nazionale di Fisica Nucleare, Sezione di Roma Tor Vergata, Rome, Italy}
\affil[3]{Università Sapienza di Roma, Rome, Italy}

\begin{document}
\maketitle
\begin{abstract}
Spiking neural networks (SNNs) run today on a fragmented landscape of hardware:
dedicated neuromorphic processors, application-specific FPGA accelerators, and
large-scale neuroscience simulators, each typically built around a fixed neuron
model, execution strategy, or workload class. We present AIGOR, a modular,
event-driven neuromorphic architecture for spiking neural network inference. AIGOR
organizes neurons into timestep-synchronized processing cores that exchange spikes
as packets over a packet-switched communication layer, and it is assembled from a
library of parameterized compute, memory, and communication IP blocks rather than
as a one-off design for a single network. The neuron model, numeric precision, the
folding of neurons onto hardware, and the partitioning across cores are configured
per instance rather than committed at design time; a single declarative
specification then generates the cores, neuron kernels, and synaptic-memory images
that realize a given network.

We validate a first prototype on the AMD Versal VPK180 across two deliberately
different workloads mapped onto the same cores: a feedforward image classifier
trained in snnTorch and a recurrent balanced random network modeled in NEST. The
classifier reproduces its snnTorch reference accuracy, and the recurrent network
matches its NEST reference at spike-level precision across multiple cores spanning
two FPGAs. We report post-implementation resource utilization and validate the
multi-node synchronization scheme in simulation up to one thousand cores on a
three-dimensional torus. The prototype's measured limits localize the throughput
bottleneck in the synaptic-delivery datapath and the global timestep barrier, and
motivate a set of datapath refinements, now in development, that the configurable
structure of the architecture admits as changes to the same cores.
\end{abstract}

\section{Introduction}
\label{sec:intro}

Spiking neural networks are studied and deployed across communities with rather
different goals and tools. In machine learning, SNNs are pursued as an
energy-efficient, event-driven alternative to conventional deep networks
\cite{roy2019towards,edgeSNN2025} and are trained with surrogate-gradient methods
\cite{neftci2019surrogate,snntorch}, often run on dedicated neuromorphic processors
\cite{truenorth,tianjic} or on
application-specific accelerators. In computational neuroscience, large recurrent networks are
simulated to study brain dynamics \cite{brunel2000} on CPU/GPU clusters or on specialized
simulation engines \cite{nest,nestgpu}. A related strand pursues SNNs for real-time inference at the
data source in high-energy-physics experiments \cite{kulkarni2023onsensor,coradin2025neuromorphic}.
Alongside these sits a broad body of FPGA implementations spanning both worlds. The
result is a fragmented landscape with little common ground across neuron models,
execution strategies, and workload classes.

Many application-specific designs commit a neuron model, a numeric precision, a
degree of parallelism, an interconnect, and a connectivity pattern at design time,
tailoring the compute logic, the software-ingestion path, and the communication to
a single use case. Such designs are effective for their target workload but are not
easily retargeted to a different network class. Rigidity is addressed in two
established ways: large programmable platforms such as SpiNNaker and Loihi run a
broad range of models on fixed hardware \cite{spinnaker,davies2018loihi,loihi2},
while a growing body of FPGA frameworks generates a configured accelerator from a
high-level description \cite{spikerplus,flexineura}. AIGOR follows the latter
direction, resolving the neuron model, numeric format, parallelism, and
partitioning when an instance is generated, so that one description can produce
hardware for different network classes.

AIGOR organizes neurons into processing cores that all advance the same discrete
timestep together, exchanging spikes as packets over a packet-switched communication layer;
input/output cores form the boundary with the outside world. The architecture is
built from modular compute, memory, and communication components, and a single
declarative specification drives a generation flow that emits a complete
synthesizable instance together with the synaptic-memory images that map a given
network onto it. The operating point of an instance, summarized by the
configuration axes in Table~\ref{tab:config-axes}, is a property of the
specification rather than a design-time commitment. The current prototype is
realized on up to two Versal VPK180 FPGAs, where the same architecture scales from
a single device to a multi-node deployment, with larger topologies exercised in
simulation.

This paper describes the architecture, the generation flow that produces instances
of it, and a methodology for validating and characterizing those instances, and it
exercises the architecture on two deliberately different SNN workload classes.
Concretely, it makes the following contributions:
\begin{itemize}
  \item \textbf{A modular, event-driven neuromorphic architecture.} AIGOR defines a
  timestep-synchronized execution model in which a library of parameterized compute,
  memory, and communication blocks composes into processing cores. Configurable
  workers host and update the neurons, accumulating each neuron's incoming synaptic
  contributions per receptor and per delay slot in circular delay buffers before the
  neuron is evaluated, and spikes travel between cores as packets over a
  packet-switched network (Section~\ref{sec:architecture}).
  \item \textbf{Configuration as an architectural property.} The neuron model,
  numeric precision, the spatial-versus-temporal folding of neurons onto hardware,
  and the partitioning of neurons across cores and workers are exposed as parameters
  of the IP blocks rather than fixed design decisions. A single declarative YAML
  specification expands together into the core RTL, the neuron kernels, and the
  initialized synaptic-memory images of a synthesizable instance
  (Section~\ref{sec:dse}).
  \item \textbf{One system across SNN regimes.} We map and run two workloads usually
  addressed by separate systems, a feedforward image classifier trained in snnTorch
  and a recurrent balanced random network (Brunel-style) modeled in NEST, onto the
  same cores, differing only in their generated configuration. The classifier
  reproduces its snnTorch reference accuracy; the recurrent benchmark is validated at
  spike-level precision against its NEST reference across multiple cores spanning two
  FPGAs (Section~\ref{sec:prelim}).
  \item \textbf{A prototype, its characterization, and a roadmap.} We report
  functional validation and post-implementation resource utilization for a Versal
  VPK180-based prototype, and validate the multi-node synchronization scheme in simulation
  up to one thousand cores. From the prototype's measured limits we derive a set of
  datapath refinements, a banked synaptic accumulator, a fused single-word spike/sync
  protocol, and globally-asynchronous locally-synchronous timestep synchronization,
  presented as the next stage of the architecture
  (Section~\ref{sec:refinements}).
\end{itemize}

AIGOR's parameterized structure means a broad range of hardware instances can be
derived from one description, providing a natural basis for the systematic
exploration of SNN hardware design trade-offs. The objective of this paper is to
present the architecture and the flow that instantiates it, to describe a prototype
on Versal FPGAs, and to validate it across distinct classes of spiking-neural-network
workloads.

\section{Background and Related Work}
\label{sec:background}

\paragraph{Spiking neural networks and their hardware.}
SNNs communicate through sparse, asynchronous spike events and carry state (membrane
potential together with synaptic and adaptation variables) that evolves in
discretized time \cite{maass1997networks,gerstner2014neuronal}. They are realized
across a wide range of hardware, from mixed-signal \cite{brainscales} and digital
neuromorphic ASICs to
FPGA accelerators and large-scale simulation engines
\cite{roy2019towards,schuman2022opportunities}. These systems differ mainly in how
much they fix in advance: application-specific accelerators commit the neuron model,
numeric precision, connectivity, and synaptic-delivery scheme up front and optimize
the result for one workload class, which is efficient for that workload but hard to
retarget.

\paragraph{Programmable neuromorphic platforms.}
A complementary approach keeps the hardware fixed and obtains breadth through
programmability. SpiNNaker realizes neuron and synapse dynamics in software on a
packet-switched array of general-purpose cores, relaxing memory coherence, synchrony,
and determinism so that arbitrary models can run on the same machine \cite{spinnaker};
its second generation keeps this software-defined approach while adding numerical
accelerators and scaling to far more cores \cite{spinnaker2}.
Intel's Loihi and Loihi~2 keep the silicon fixed while exposing programmable neuron
models, graded spikes, and on-chip learning over an asynchronous mesh
\cite{davies2018loihi,loihi2}. These designs show that a neuromorphic architecture
can be valued for the range of models and workloads it supports rather than for a
single efficiency figure. AIGOR shares this orientation but reaches it differently:
the model, precision, datapath, and parallelism are resolved when an instance is
generated and synthesized, so each configuration is a tailored hardware build rather
than a program running on fixed silicon.

\paragraph{FPGA SNN systems.}
On FPGAs, two tendencies are visible. One targets a specific workload at scale:
NeuroAIx, for example, is an FPGA cluster for accelerated, deterministic neuroscience
simulation, built around leaky integrate-and-fire neurons with current-based synapses
for the cortical-microcircuit benchmark \cite{neuroaix}. The other generates a
configured accelerator from a high-level description, as in Spiker+, which produces
compact multi-layer SNN accelerators for edge inference \cite{spikerplus}; recent work
continues in this direction, adding configurable precision and topology or multi-FPGA
scaling \cite{flexineura,neuroring,neurocorex}. AIGOR belongs to this second group and
differs in the range of choices it draws from one specification: neuron model, numeric
precision, spatial-temporal folding, and partitioning across cores. Where these systems
usually hold one of these fixed, a neuron model or a workload class, AIGOR leaves them
open and exercises both a feedforward classifier and a recurrent network on the same
cores.

\paragraph{Composition and transport.}
AIGOR builds its inter-core communication on the \textsc{Apeiron}
framework~\cite{apeiron}, a low-latency, packet-switched fabric developed
for distributed real-time inference in high-energy-physics data acquisition, a domain with a
long line of FPGA-based front-end and inference systems
\cite{felix,Duarte:2018ite,Aarrestad_2021,Perticaroli:fpgarich}. This
lineage orients AIGOR toward streaming inference at the data source, a setting we
pursue for the ePIC dual-radiator RICH detector \cite{VALLARINO2024168834} in a companion
study \cite{aigor_drich}, and provides the multi-node transport over which a
partitioned network spans several FPGAs.

\section{AIGOR Architecture}
\label{sec:architecture}

This section describes the AIGOR architecture as realized in the current prototype;
the generation flow that produces a concrete instance is described in
Section~\ref{sec:dse}. The structural choices below, the neuron model, the numeric
format, the datapath width, the degree of parallelism, and the partitioning of neurons
across cores and workers, are set when an instance is generated rather than fixed once
for all instances; together they form the configuration axes of
Table~\ref{tab:config-axes}. All instances share the same interfaces, the same
timestep-synchronized execution, and the same routing fabric, so that a given instance
is one point in this space.

\begin{table}[htbp]
  \centering
  \caption{Principal AIGOR configuration axes exposed to design-space exploration in
  the current prototype. Additional axes introduced by the refinements of
  Section~\ref{sec:refinements} (synaptic-delivery scheme and inter-core routing
  policy) extend this space.}
  \label{tab:config-axes}
  \small
  \begin{tabular}{@{}lll@{}}
    \toprule
    \textbf{Axis} & \textbf{Variants / range} & \textbf{Trade-off governed} \\
    \midrule
    Numeric format     & float / fixed-point (tunable widths) & accuracy vs.\ area/timing \\
    Neuron model       & LIF / IAF / adaptive-exp.\ / \dots   & expressivity vs.\ cost \\
    Neuron compute     & spatial / time-multiplexed           & latency vs.\ area \\
    Cores per system   & $\geq 1$ (multi-FPGA)                & capacity / partitioning \\
    Workers per core   & $\geq 1$                             & intra-core parallelism \\
    Neurons per worker & $\geq 1$                             & spatial--temporal folding \\
    \bottomrule
  \end{tabular}
\end{table}

\subsection{System organization}
\label{sec:arch:system}
\begin{figure}[htbp]
  \centering
  \includegraphics[width=\linewidth]{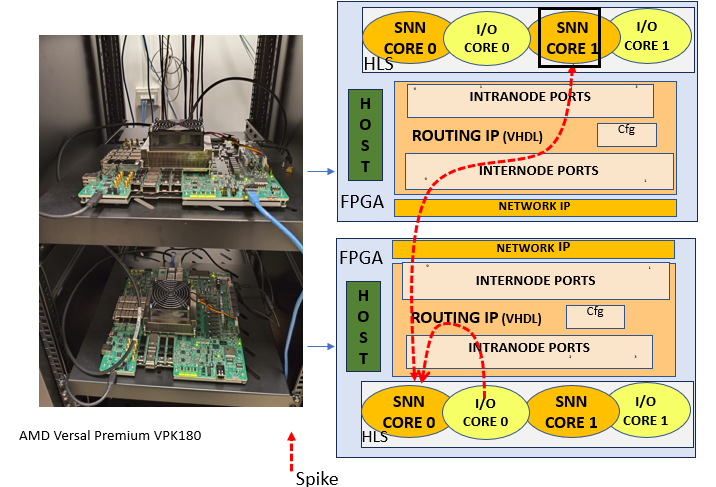}
  \caption{AIGOR prototype deployed across two AMD Versal Premium VPK180 boards.
  \emph{Left:} the two-board setup. \emph{Right:} the per-FPGA organization. On each
  device the SNN cores and I/O cores are synthesized as HLS kernels and attach to the
  \textsc{Apeiron} routing IP (VHDL), which exposes \emph{intra-node} ports for on-chip
  core-to-core traffic and \emph{inter-node} ports that reach the network IP driving the
  physical inter-FPGA link; a configuration block (Cfg) holds the routing tables, and the
  host (the board's Arm APU) injects stimulus and reads results through the I/O cores. The
  red dashed path traces a single spike: originating on the lower FPGA, it is switched by
  the routing IP and, because its postsynaptic target resides on the other device, leaves
  through the inter-node ports and network IP, crosses to the upper FPGA, and is delivered
  by that FPGA's routing IP to the destination SNN core (highlighted). Delivery is
  transparent to the cores: the same packet interface carries a spike whether its target
  sits on the same device or across the torus.}
  \label{fig:sys-overview}
\end{figure}

An AIGOR system is a set of \emph{cores} interconnected by a routing fabric (Fig.~\ref{fig:sys-overview}). 
\emph{SNN
cores} host the neurons and run their dynamics; \emph{I/O cores} form the boundary with
the outside world, encoding external data (detector frames, event-camera streams, or
externally generated Poisson sources) into spikes and collecting results. In the
current prototype a \emph{host}, the Versal board's Arm-based Application Processing
Unit (APU), injects stimulus and reads outputs through the I/O cores, and reads and
writes status and configuration registers. Cores occupy positions on a logical
three-dimensional torus of nodes, so that a single specification scales from one device
to a multi-FPGA partition without changing the core design.

Inter-core transport is provided by the \textsc{Apeiron} framework
(Section~\ref{sec:background}), which supplies a packet-switched, low-latency
communication IP, the \emph{routing IP}, that carries events between cores intra-node
and inter-node across the torus. Each AIGOR core attaches to this routing IP and
exchanges spikes as packets. Because computation and transport are decoupled, the cores
are composable: the routing IP is the same fabric in every configuration.

Execution combines \emph{event-driven} communication with global \emph{timestep
synchronization}. Communication and synaptic gathering are event-driven: only the spikes
actually emitted in a timestep trigger synaptic-memory fanout reads and accumulation, so
a core does no synaptic work for silent inputs. Neuron updates, by contrast, are
timestep-driven: every neuron's state is advanced once per timestep, so the model is
discrete-time. Within a timestep, the spikes emitted in the previous step are delivered,
accumulated, and integrated; the step then closes with a synchronization (\emph{sync})
event that advances each core's timestep counter and keeps all cores globally
synchronized, ensuring no spike is misordered or lost. Every event a core exchanges
carries a leading sync bit: a spike event carries the presynaptic identifier in the
manner of an address-event representation \cite{AER:ref}, while a
sync event carries only the originating core and the timestep, signaling the barrier. In
the prototype the barrier is a dedicated per-core sync packet, exchanged once per
timestep. Sample boundaries, when required for inference, are handled either by an
end-of-sample marker or by a fixed timestep window driven from a configuration register,
at which point the neuron states are reset.

\subsection{The SNN core}
\label{sec:arch:core}
A network is partitioned hierarchically by the \emph{AIGOR loader}, a software toolchain
that groups neurons into workers, workers into cores, and cores across FPGAs, and emits
the per-core synaptic-memory images that encode the mapped connectivity. Two configurable
choices determine how a core realizes its neurons: how each worker computes its neurons,
and how synaptic input reaches them.

The first choice is how a worker computes its neurons. In the \emph{spatial}
(worker-parallel) organization a worker holds one physical datapath per neuron and updates
them concurrently, maximizing parallelism and minimizing latency; in the
\emph{time-multiplexed} organization a single datapath is folded over the
worker's neurons, exchanging cycles for area and making large populations tractable on a
fixed silicon budget. Combined with the partitioning knobs (number of cores, workers per
core, and neurons per worker) this places the same network anywhere from maximally
parallel to maximally folded.

\begin{figure}[htbp]
  \centering
  \includegraphics[width=1.2\linewidth]{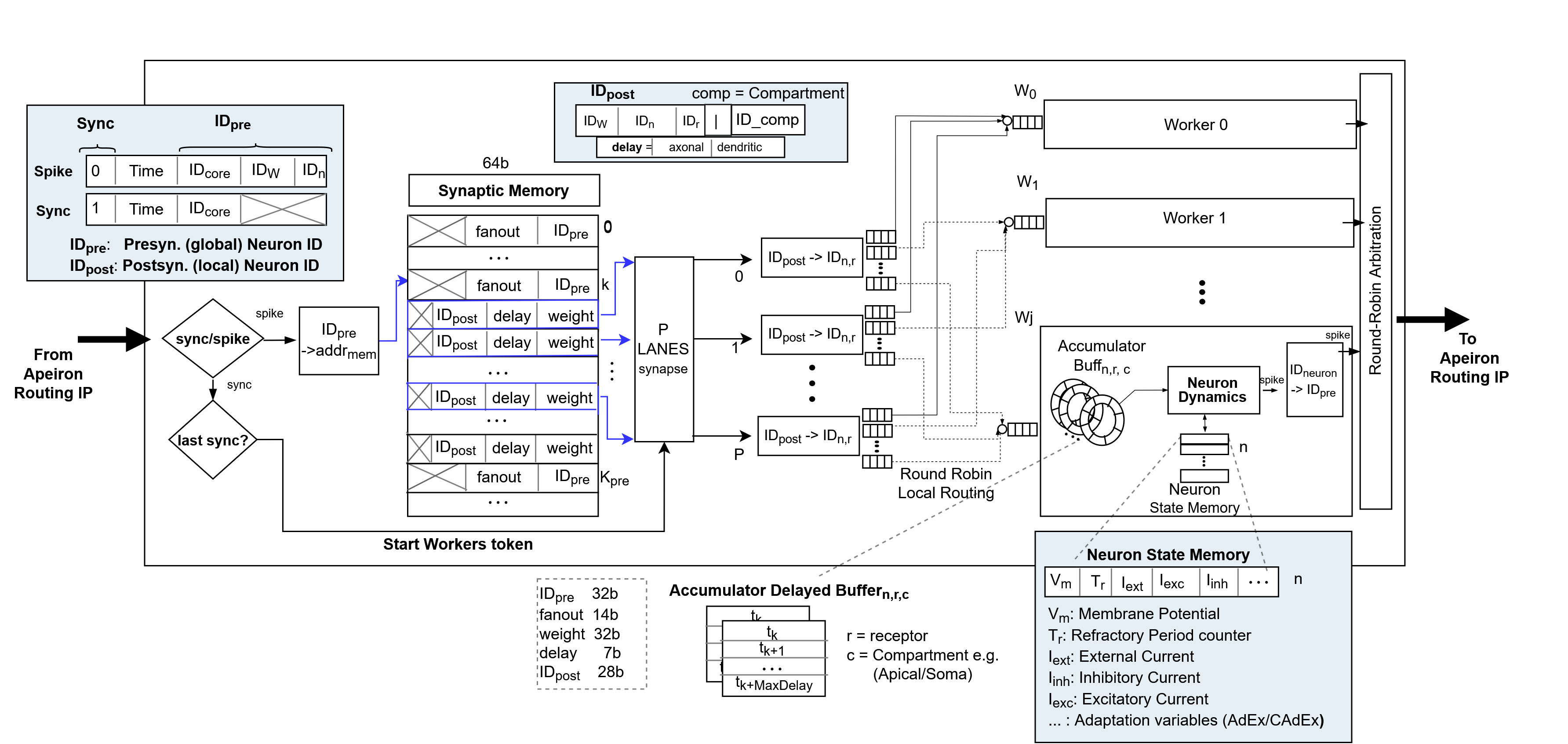}
  \caption{Prototype SNN core with local per-worker synaptic routing. Events arrive from
  the \textsc{Apeiron} communication IP carrying a leading sync bit: a spike event carries
  the presynaptic identifier
  ($\text{ID}_{\text{pre}}=\langle\text{ID}_{\text{core}},\text{ID}_W,\text{ID}_n\rangle$),
  a sync event only the originating core and timestep. A spike indexes the synaptic memory;
  the fetched fanout (weight, delay, receptor, destination $\text{ID}_{\text{post}}$) is
  distributed to the destination workers, resolved to local (neuron, receptor) pairs,
  buffered in per-receptor FIFOs, and merged by a round-robin stage. Each worker accumulates
  contributions into per-(neuron, receptor) circular delay buffers ($\text{Buffer}_{n,r}$),
  updates its neurons against the neuron-state memory, and emits spikes that are re-encoded
  and arbitrated back onto the fabric.}
  \label{fig:proto-core}
\end{figure}

The second choice is how synaptic input is delivered. The prototype uses \emph{local
per-worker routing} (Fig.~\ref{fig:proto-core}), in which accumulation happens inside the
workers rather than in a shared block. An incoming spike indexes a \emph{synaptic memory},
held in on-chip URAM on the VPK180, that stores per presynaptic source the fanout of
target connections. A \emph{memory handler} streams the addressed fanout and unpacks each
connection into a synapse word carrying its weight, synaptic delay, and receptor
identifier, marking the last word of the timestep's traffic so downstream stages know when
the barrier has been reached. Each synapse word is then routed to the worker that hosts its
destination neuron, where the destination identifier $\text{ID}_{\text{post}}$ is resolved
into a local (neuron, receptor) pair. Routed words are buffered in per-receptor input FIFOs
and merged by a round-robin stage into each worker's accumulation path. There, each
contribution lands in a circular delay buffer held per (neuron, receptor): a
$\texttt{MAX\_DELAY}$-slot ring indexed by the synaptic delay relative to a head pointer
that rotates once per timestep, so that a delayed contribution surfaces in the correct
future timestep. At the timestep boundary each buffer's head slot is read into the worker's
neuron-dynamics stage and cleared. Neuron state (membrane potential, refractory counter,
per-receptor currents, and any adaptation variables) lives in a per-worker neuron-state
memory. Spikes emitted on firing are re-encoded to the global presynaptic identifier, merged
by a round-robin arbiter, and returned to the \textsc{Apeiron} fabric, where they are routed
to the cores hosting their postsynaptic targets, closing the loop.

The data plane is $P$ words wide, where $P$ is a configurable number of parallel
synaptic-memory read lanes, and stages communicate through valid/ready handshakes with
per-neuron buffering, so that the serial readout of accumulated input overlaps the workers'
independent neuron updates. This delivery path scales with the number of workers but
concentrates two costs that the prototype's measurements reveal (Section~\ref{sec:prelim}):
the synaptic-memory fetch, which dominates for densely-connected layers because every
incoming spike retrieves its target population's entire fanout, and the per-worker routing
and round-robin merge, whose serialization grows with the spike load per timestep.
Section~\ref{sec:refinements} describes a redesign of this stage, a banked synaptic
accumulator with independent read-modify-write lanes, that targets both.

\subsection{Neuron model and numeric format}
\label{sec:arch:model}
The neuron dynamics are a replaceable HLS kernel: leaky integrate-and-fire,
integrate-and-fire, and adaptive-exponential models are currently supported. The numeric
representation is independently selectable between floating point and fixed point with
tunable field widths, exposing the accuracy-versus-area/timing trade-off without touching
the surrounding datapath.

\section{Configuration and Generation Flow}
\label{sec:dse}

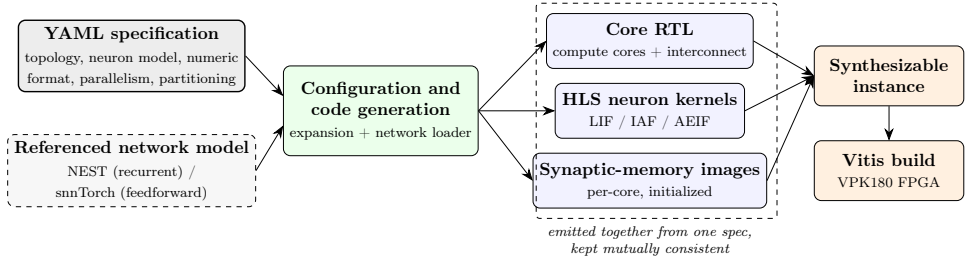
\begin{figure}[htbp]
\centering
\resizebox{\linewidth}{!}{%
\begin{tikzpicture}[
  >={Stealth[length=2.4mm]}, font=\small,
  src/.style={draw,thick,rounded corners,align=center,fill=black!7,
              minimum width=34mm,minimum height=13mm},
  ref/.style={draw,dashed,rounded corners,align=center,fill=black!3,
              minimum width=34mm,minimum height=12mm},
  gen/.style={draw,rounded corners,align=center,fill=green!8,
              minimum width=28mm,minimum height=18mm},
  art/.style={draw,rounded corners,align=center,fill=blue!5,
              minimum width=38mm,minimum height=11mm},
  inst/.style={draw,rounded corners,align=center,fill=orange!12,
               minimum width=30mm,minimum height=12mm},
  note/.style={font={\itshape\scriptsize},align=center}]

  \node[src] (yaml) at (0,0)
    {\textbf{YAML specification}\\[1pt]{\scriptsize topology, neuron model, numeric}\\{\scriptsize format, parallelism, partitioning}};
  \node[ref] (net) at (0,-2.3)
    {\textbf{Referenced network model}\\[1pt]{\scriptsize NEST (recurrent) /}\\{\scriptsize snnTorch (feedforward)}};

  \node[gen] (gen) at (5.0,-1.1)
    {\textbf{Configuration and}\\\textbf{code generation}\\[1pt]{\scriptsize expansion + network loader}};

  \node[art] (rtl) at (10.4,0.3)
    {\textbf{Core RTL}\\{\scriptsize compute cores + interconnect}};
  \node[art] (ker) at (10.4,-1.1)
    {\textbf{HLS neuron kernels}\\{\scriptsize LIF / IAF / AEIF}};
  \node[art] (mem) at (10.4,-2.5)
    {\textbf{Synaptic-memory images}\\{\scriptsize per-core, initialized}};

  \node[inst] (inst) at (15.2,-0.4)
    {\textbf{Synthesizable}\\\textbf{instance}};
  \node[inst] (fpga) at (15.2,-2.3)
    {\textbf{Vitis build}\\{\scriptsize VPK180 FPGA}};

  \draw[->] (yaml.east) -- ([yshift=3mm]gen.west);
  \draw[->] (net.east)  -- ([yshift=-3mm]gen.west);
  \draw[->] (gen.east) -- (rtl.west);
  \draw[->] (gen.east) -- (ker.west);
  \draw[->] (gen.east) -- (mem.west);
  \draw[->] (rtl.east) -- (inst.west);
  \draw[->] (ker.east) -- (inst.west);
  \draw[->] (mem.east) -- (inst.west);
  \draw[->] (inst.south) -- (fpga.north);

  \draw[dashed,rounded corners=2pt]
    ($(rtl.north west)+(-2mm,2mm)$) rectangle ($(mem.south east)+(2mm,-2mm)$);
  \node[note] at ($(mem.south)+(0,-6.5mm)$)
    {emitted together from one spec,\\kept mutually consistent};
\end{tikzpicture}%
}
\caption{AIGOR configuration-driven generation flow. A single YAML specification is
expanded by the code-generation stage into the core RTL, the HLS neuron kernels, and the
initialized synaptic-memory images, which together form a synthesizable instance for the
target FPGA platform.}
\label{fig:toolflow}
\end{figure}

An AIGOR instance is produced by an automated generation flow driven from a single
declarative specification, rather than hand-specialized per network
(Fig.~\ref{fig:toolflow}). All structural choices, network topology, neuron model, numeric
format, the degree of spatial versus temporal parallelism, and the partitioning of neurons
across cores and workers, are captured in one YAML specification. A configuration stage
expands it into the three artifacts that constitute an instance: the register-transfer
description of the compute cores and their interconnect, the HLS neuron kernels, and the
initialized synaptic-memory images that encode the target network's connectivity. Because
every artifact is emitted from the same specification, an architectural variant is a
one-line change rather than a manual rewiring across heterogeneous source files, which keeps
systematic, reproducible exploration practical.

The processing cores, the workers, handlers, and the control and routing that ties them
together, are authored as a parameterized cycle-accurate SystemC model and lowered to
synthesizable RTL by Catapult HLS. The boundary and communication kernels (the I/O cores)
are written as Vitis HLS tasks against the \textsc{Apeiron} API, and the most
latency-sensitive synaptic datapath, the memory handler, is implemented in VHDL. These three
paradigms are unified by the shared specification and build system, which emits the SystemC
parameters, the VHDL configuration package, and the HLS task wiring.

The same SystemC description doubles as a fast functional simulator. It provides
pre-synthesis validation of a configured instance, checking its numerical behavior and
exercising the spike/sync protocol without committing to a full hardware build, and a
complementary transaction-level (TLM) model of the routing IP lets multi-node topologies be
simulated well beyond the available FPGA count; in this way the multi-node communication IP
and the barrier-based timestep synchronization were exercised up to 1000 cores on a
$10\times10\times10$ torus, confirming the synchronization scheme at a scale beyond the
available hardware. Networks enter the flow through a common ingestion path:
recurrent models built in NEST~\cite{nest} and feedforward models trained in
snnTorch~\cite{snntorch} are exported to a
common intermediate representation, from which the toolchain derives the per-core
synaptic-memory images. A single platform thus serves both the machine-learning-oriented and
the neuroscience-oriented workloads usually addressed by separate tools.

\section{Methodology and Benchmarks}
\label{sec:methodology}

We evaluate AIGOR with two contrasting workloads and characterize the configured instances
for functional correctness and resource utilization. All hardware results are obtained on AMD
Versal VPK180 devices (one or two boards depending on the benchmark). Unless otherwise stated,
the datapath uses 32-bit fixed-point arithmetic with 12 integer bits.

\subsection{Benchmarks}
\label{sec:meth:bench}
Both workloads use a leaky integrate-and-fire neuron, but each matches the conventions of its
reference tool: the feedforward classifier follows the snnTorch \texttt{Leaky} neuron, the
recurrent network the NEST \texttt{iaf\_psc\_delta} model. The two agree in their integration
core and differ in minor details of reset and input handling, which the corresponding kernel
configuration reproduces.
\paragraph{Feedforward: spiking MNIST.} A fully-connected spiking classifier trained in
snnTorch~\cite{snntorch}, $256\to128\to10$ leaky integrate-and-fire neurons, with
$28\times28$ MNIST images downsampled on-chip to $16\times16$ by the input core and
rate-encoded over a 25-timestep window. Both layers are placed on a single core in the
spatial (worker-parallel) organization of Section~\ref{sec:arch:core}, where each neuron has
its own datapath and the timestep dynamics of all the core's neurons are computed in parallel
(8 workers, 32 neurons each).

\paragraph{Recurrent: balanced random network.} A Brunel-style~\cite{brunel2000}
sparsely-connected excitatory/inhibitory network modeled in NEST~\cite{nest} ($\sim$1638
excitatory / 410 inhibitory neurons, connection probability $0.1$, driven by an $800$~Hz
Poisson source). This workload
runs across multiple cores on two FPGAs and validates AIGOR on cyclic connectivity. Both
networks are exported to the common intermediate representation and mapped onto the same
architecture, differing only in configuration.

\subsection{Functional validation and precision}
\label{sec:meth:func}
For each benchmark we establish correctness against the software reference by deploying the
configured instance on the FPGA and comparing its output under stimuli identical to the
reference. The recurrent network is compared to its NEST reference at the spike level
(matching spike times, or agreement of population statistics where fixed-point rounding
accumulates); the classifier is compared to its snnTorch reference by classification
accuracy. The same cycle-accurate SystemC model used during development provides an
intermediate, pre-synthesis check. We assess the effect of fixed-point precision by comparing
the hardware spike output against the reference at a representative fixed-point setting and
establishing the regime over which the two agree exactly; a full sweep across field widths is
part of the planned characterization of Section~\ref{sec:refinements}.

\section{Results}
\label{sec:prelim}

We report a first characterization of the prototype: functional validation on both workload
classes, resource utilization on the VPK180, and validation of the multi-node
synchronization scheme in simulation. The measured throughput of the prototype localizes the
bottleneck that motivates the refinements of Section~\ref{sec:refinements}.

\subsection{Functional correctness}
\label{sec:res:func}
\begin{figure}[htbp]
  \centering
  \includegraphics[width=\linewidth]{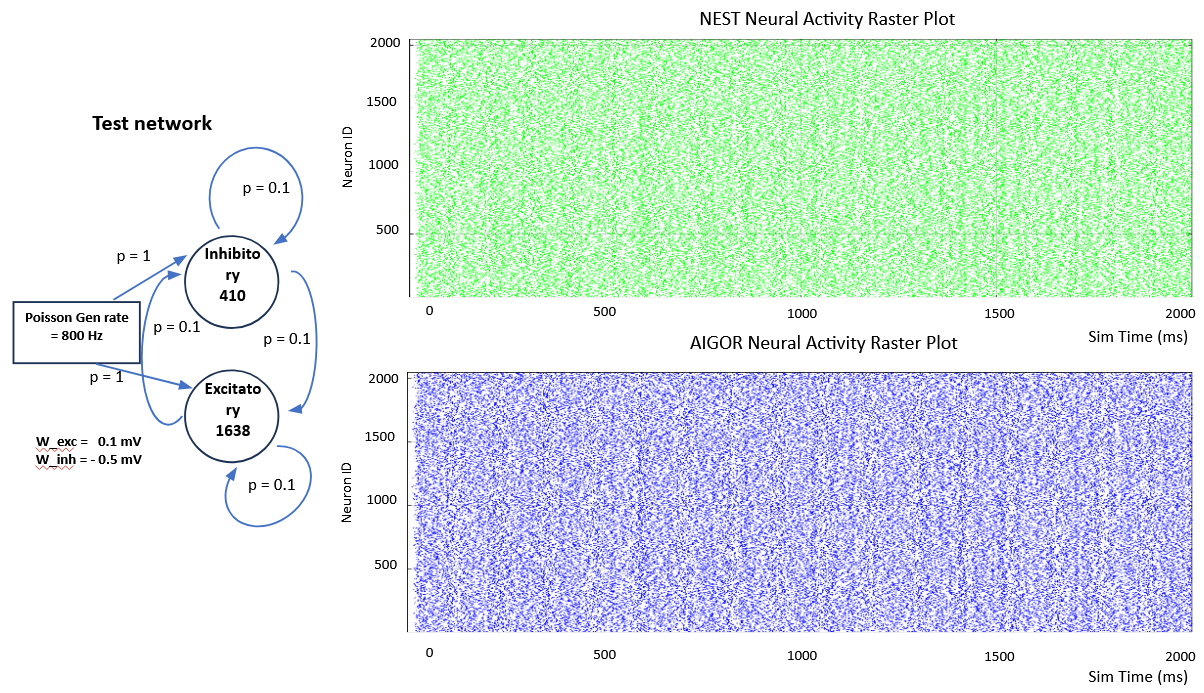}
  \caption{Functional validation on the recurrent benchmark. \emph{Left:} the balanced random test network, a Brunel-style excitatory/inhibitory population of 1638 excitatory and 410 inhibitory leaky integrate-and-fire neurons with recurrent connection probability $p=0.1$ and synaptic weights $W_{\text{exc}}=0.1$~mV, $W_{\text{inh}}=-0.5$~mV, driven by an $800$~Hz Poisson source connected with $p=1$. \emph{Right:} spike raster of all 2048 neurons over the simulated interval (0.1~ms timestep) for the NEST reference (top) and the AIGOR FPGA instance (bottom). Both settle into the same asynchronous-irregular regime, and
  their population firing statistics agree across the window; exact per-spike agreement holds over the initial timesteps before fixed-point rounding diverges the individual neuron trajectories.}
  \label{fig:nest_brunel}
\end{figure}

On the feedforward benchmark, the deployed classifier reproduces the classification accuracy
of its snnTorch reference ($\sim$95\% on the MNIST test set) at the 25-timestep encoding
window. On the recurrent benchmark, the FPGA reproduces the firing pattern of the NEST
reference over the exercised interval (up to 3~ms of simulated time at a 0.1~ms timestep), on
up to 4 cores on a single FPGA and up to 8 cores across two FPGAs. Under the representative
fixed-point setting (32 total bits, 12 integer), the hardware reproduces the reference spike
times \emph{exactly} for up to $\sim$2~ms of simulated time \ref{fig:nest_brunel}; beyond that, fixed-point rounding
gradually accumulates.

\subsection{Throughput}
\label{sec:res:tput}
On the classifier, over 10{,}000 MNIST samples, the single-core instance sustains 568
samples/s, measured with an on-chip cycle counter. Throughput is set by the spike load per
timestep, and hence by the input encoding and connection density: for a fully-connected
network at this scale the synaptic-memory fetch dominates, since each incoming spike retrieves
its target population's entire fanout. This is the primary bottleneck that the banked
accumulator of Section~\ref{sec:refinements} is designed to relieve.

\subsection{Resource utilization}
\label{sec:res:util}
Table~\ref{tab:util} reports post-implementation utilization on the VPK180 for a single core
with the time-multiplexed worker, swept across workers ($W$) and neurons per worker ($n$).
Utilization is LUT-dominated and scales with the degree of spatial parallelism; at 2048
neurons a single core uses of order one fifth of the device LUTs and a small fraction of its
BRAM, with the synaptic image held on-chip. Scaling a single core beyond this was limited by
place-and-route rather than by logic or memory capacity.

\begin{table}[htbp]
  \centering
  \caption{Post-implementation utilization on the VPK180 (Total PL memory 994~Mb) for a
  single core with the time-multiplexed worker, as a function of workers ($W$) and neurons
  per worker ($n$).}
  \label{tab:util}
  \begin{tabular}{@{}lrrrrr@{}}
    \toprule
    \textbf{Config ($W\times n$)} & \textbf{LUT} & \textbf{FF} & \textbf{DSP} & \textbf{BRAM} & \textbf{Syn.\ mem} \\
    \midrule
    16 n\ \ ($4\times4$)      & 0.40\% & 0.32\% & 0  & 0.04\% & 16~Kb  \\
    64 n\ \ ($4\times16$)     & 0.63\% & 0.47\% & 0  & 0.08\% & 197~Kb \\
    512 n\ \ ($4\times128$)   & 2.25\% & 0.45\% & 0  & 0.32\% & 4~Mb   \\
    2048 n\ ($32\times64$)    & 22.7\% & 2.93\% & 2  & 1.94\% & 84~Mb  \\
    2048 n\ ($16\times128$)   & 21.1\% & 1.59\% & 34 & 1.30\% & 84~Mb  \\
    \bottomrule
  \end{tabular}
\end{table}

The worker-parallel organization trades on-chip memory for latency and throughput, and its
two memories scale differently. The neuron-state memory grows linearly with the neuron count:
in the spatial organization every neuron datapath must read and write its own state in the
same cycle, so each neuron's state occupies an independent memory block, and a single core of
$\sim$2k such neurons uses of order 33\% of the device BRAM on state alone. Because a BRAM
block exposes two ports, one block can back two neurons before a third would contend;
time-multiplexing a group of neurons onto one datapath removes the per-neuron block at the
cost of serialized updates, and is the lever used to place larger populations on a fixed
budget. The synaptic memory, by contrast, grows with the connection count: for a
fully-connected layer it scales quadratically with the neuron count, so the same $\sim$2k-neuron
core carries $\sim$4M synapses. Held in URAM this reaches $\sim$38\% of the device for this
configuration and stays on-chip; beyond the on-chip capacity the synaptic image would have to
move to off-chip DRAM, whose higher and less predictable access latency is especially costly
for sparse SNNs, where the fanout reads are short and irregular rather than long bursts.

\section{Architectural Refinements and Future Work}
\label{sec:refinements}

The prototype's measured throughput points to two limits of the current core: the
synaptic-delivery datapath and the global timestep barrier. This section sets out the
refinements that address them. Each is an incremental addition to the same core rather than a
new design, specified here ahead of implementation and on-FPGA measurement, together with the
characterization we plan to run on the resulting instances.

\subsection{Banked synaptic accumulator}
\label{sec:ref:accum}
\begin{figure}[htbp]
\centering
\begin{subfigure}[t]{0.48\textwidth}
    \centering
    \includegraphics[width=\linewidth]{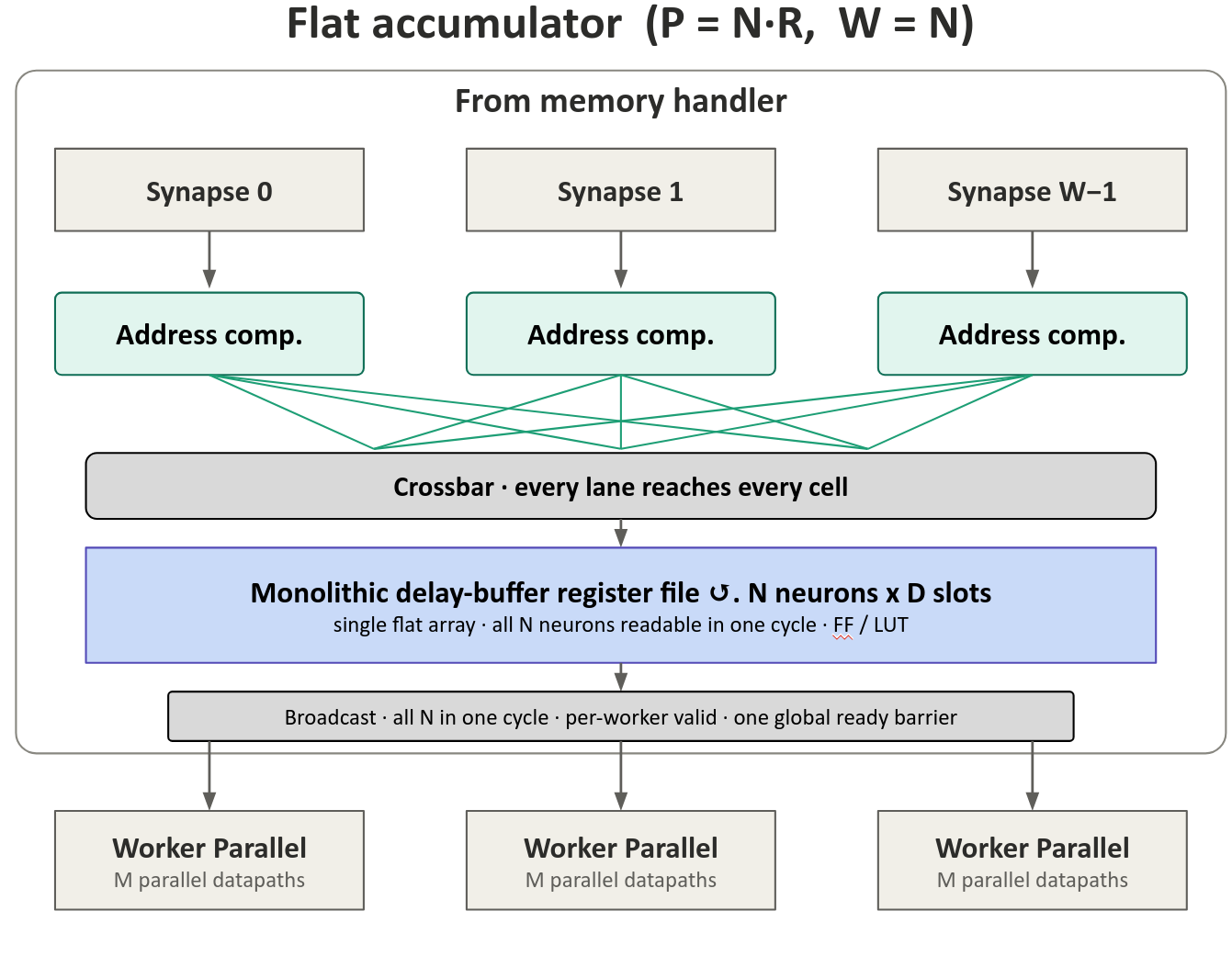}
    \caption{Flat accumulator: fully parallel, single delay buffer, global synchronization.}
    \label{fig:variants-flat}
\end{subfigure}
\hfill
\begin{subfigure}[t]{0.48\textwidth}
    \centering
    \includegraphics[width=\linewidth]{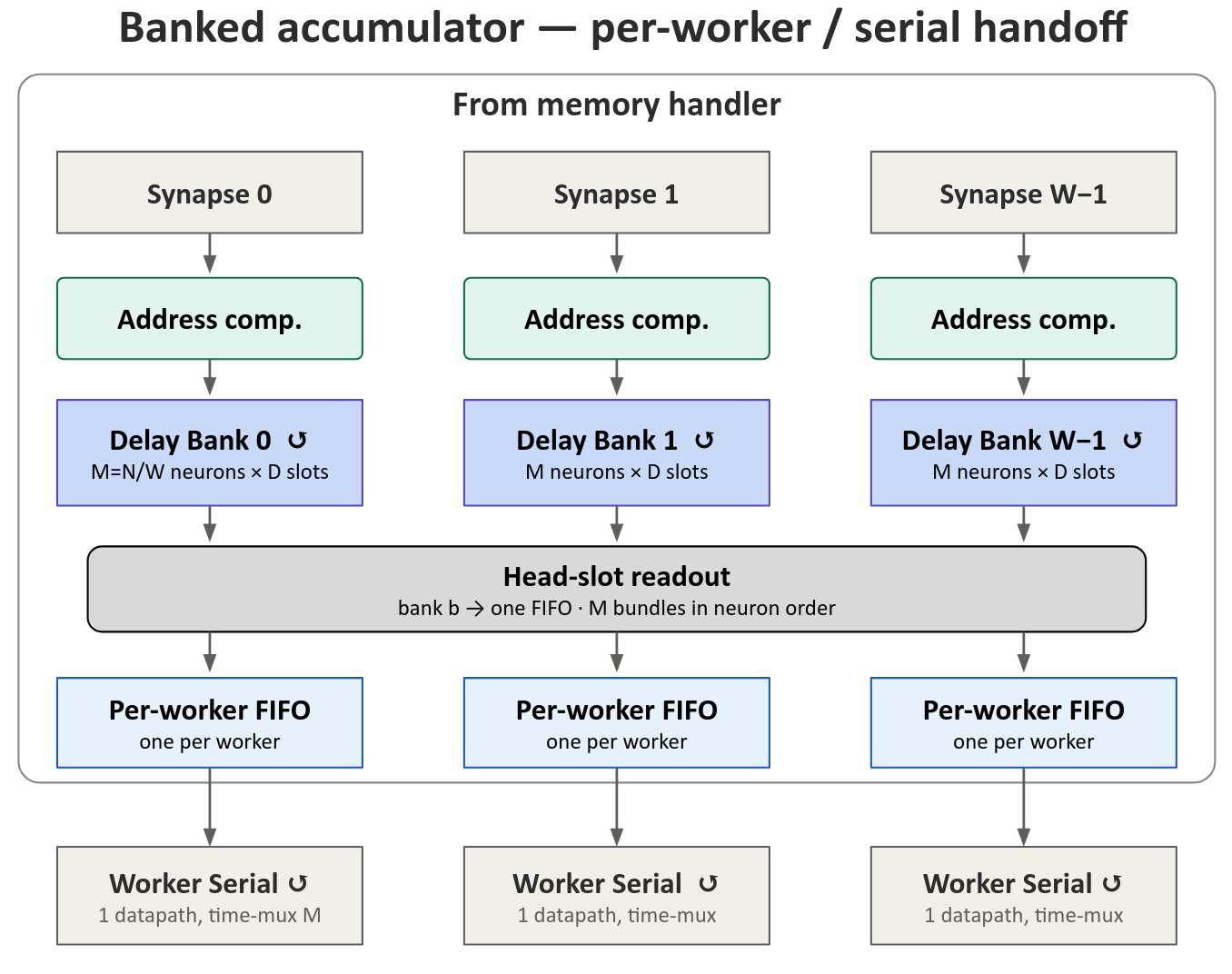}
    \caption{Banked accumulator: per-worker banking with streaming readout and
    time-multiplexed execution.}
    \label{fig:variants-worker}
\end{subfigure}

\vspace{1em}

\begin{subfigure}[t]{0.60\textwidth}
    \centering
    \includegraphics[width=\linewidth]{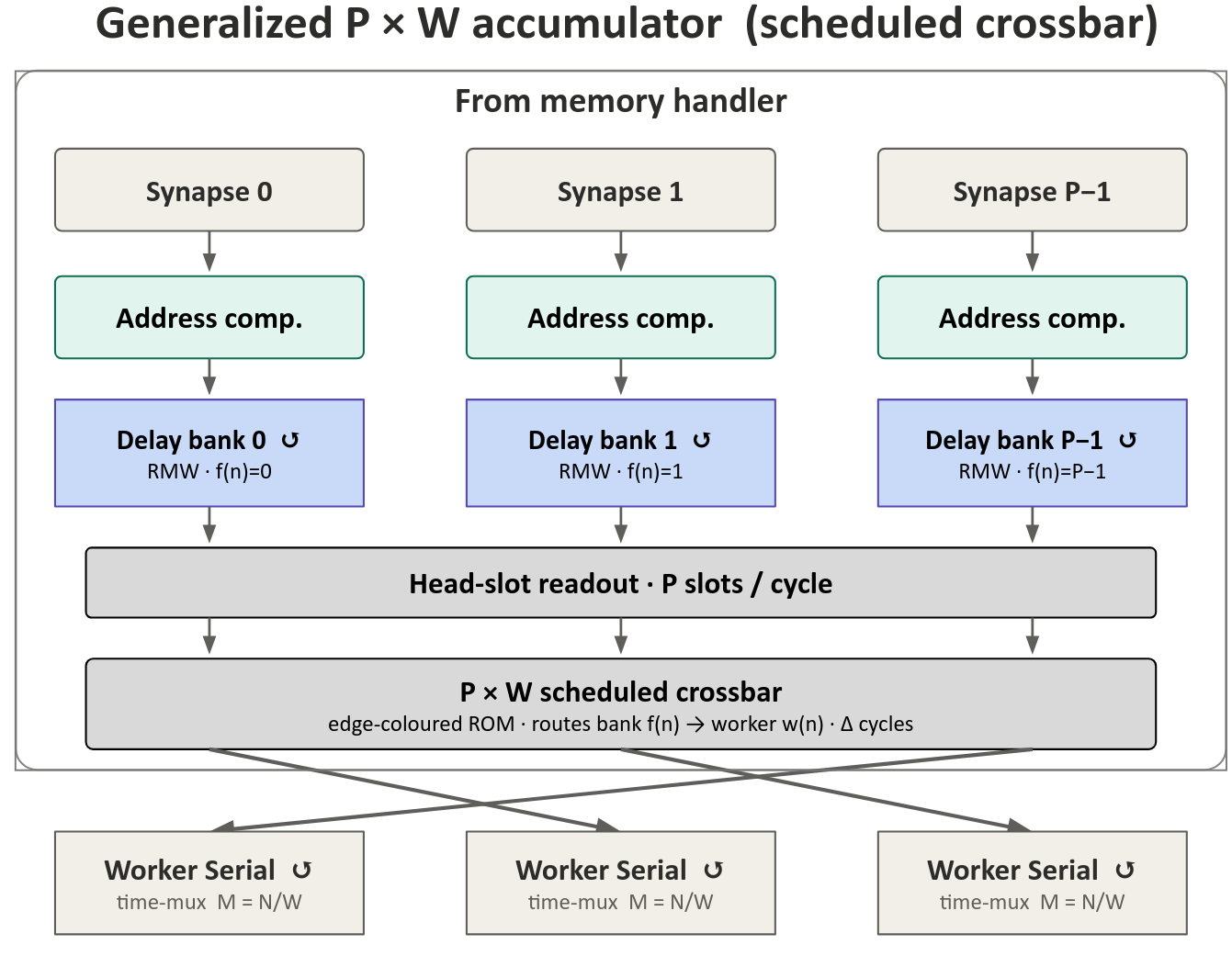}
    \caption{Generalized $P\times W$ banked accumulator with scheduled crossbar.}
    \label{fig:variants-new}
\end{subfigure}

\caption{Design space for the banked synaptic-accumulator redesign. \emph{Flat accumulator}
(fully parallel): a crossbar matches every input lane to every cell of a single flat
register-file delay buffer, which is read in one cycle and broadcast to all workers behind one
global ready barrier; each worker steps its $M$ neuron datapaths together. \emph{Banked
accumulator}, per-worker / serial handoff ($P = W$, $f$ aligned to $w$): $P$ addressable-RAM
banks perform read-modify-write accumulation; the readout streams bank $b$'s $M$ cells in order
into one per-worker FIFO, and worker $b$ time-multiplexes its $M = N/W$ neurons over $M$ cycles,
rate-matched to the readout. \emph{Generalized $P\times W$ accumulator} (scheduled-crossbar
handoff): $f(n)$ maps a neuron to its bank, $w(n)$ to its worker, with $f(n)\neq w(n)$ in
general; writes are bucketed by $f(n)$ (no input crossbar) with compile-time pre-resolved
bank-local addresses, and at the barrier a $P\times W$ crossbar, configured per cycle by a
compile-time edge-colouring, routes each bank's head-slot stream to worker $w(n)$ in $\Delta$
cycles (crossing lines show one edge-colour). Throughout, a cell is one neuron's circular delay
line within its bank, one \texttt{MAX\_DELAY}-slot buffer holding $R$ receptor accumulators per
slot.}
\label{fig:variants}
\end{figure}

The prototype's per-worker routing and round-robin merge (Section~\ref{sec:arch:core})
serialize with the spike load. The proposed replacement moves accumulation into a shared bank
of read-modify-write lanes: the neurons of a core are split into $P$ banks updated in parallel
by $P$ lanes, one lane per bank, with no input crossbar or arbitration on the delivery path.
The loader pre-sorts each presynaptic source's fanout into $P$ per-bank buckets, padded to a
common length so the lanes advance together; the padding is small for the connectivity of
interest, since a fully-connected layer splits evenly across banks and sparse random
connectivity is balanced in expectation. Within a bank, each neuron's pending input occupies a
circular delay buffer addressed as
$\text{cell}\cdot\texttt{MAX\_DELAY}+((\text{head}+\delta)\bmod\texttt{MAX\_DELAY})$, with a
write-first read and a sticky bypass register keeping repeated accumulation into the same cell
coherent. Figure~\ref{fig:variants} lays out the design space this admits, from a fully-parallel
flat point, through a per-worker banked point ($P=W$, banks aligned to workers), to a
generalized $P\times W$ point with a scheduled crossbar.

\subsection{Inter-core routing topology}
\label{sec:ref:routing}
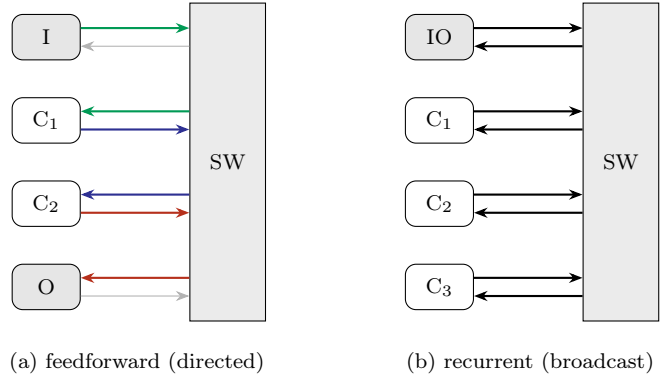
\begin{figure}[htbp]
  \centering
  \begin{tikzpicture}[
      core/.style={draw,rounded corners,minimum width=9mm,minimum height=6mm,font=\scriptsize},
      io/.style={draw,rounded corners,fill=black!10,minimum width=9mm,minimum height=6mm,font=\scriptsize},
      sw/.style={draw,fill=black!8,minimum width=10mm,minimum height=42mm,font=\scriptsize},
      >={Stealth[length=1.8mm]},
      idle/.style={draw=black!30},
      every node/.style={inner sep=2pt}]

    \begin{scope}
      \node[io]   (fi)  at (0,0)      {I};
      \node[core] (fc1) at (0,-1.1)   {C$_1$};
      \node[core] (fc2) at (0,-2.2)   {C$_2$};
      \node[io]   (fo)  at (0,-3.3)   {O};
      \node[sw]   (fs)  at (2.4,-1.65) {SW};

      \draw[->,ForestGreen,thick] ([yshift=1.2mm]fi.east) -- ([yshift=1.2mm]fi.east -| fs.west);
      \draw[->,idle]              ([yshift=-1.2mm]fi.east -| fs.west) -- ([yshift=-1.2mm]fi.east);
      
      \draw[->,ForestGreen,thick] ([yshift=1.2mm]fc1.east -| fs.west) -- ([yshift=1.2mm]fc1.east);
      \draw[->,Blue,thick]    ([yshift=-1.2mm]fc1.east) -- ([yshift=-1.2mm]fc1.east -| fs.west);
      
      \draw[->,Blue,thick]    ([yshift=1.2mm]fc2.east -| fs.west) -- ([yshift=1.2mm]fc2.east);
      \draw[->,BrickRed,thick]    ([yshift=-1.2mm]fc2.east) -- ([yshift=-1.2mm]fc2.east -| fs.west);
      
      \draw[->,BrickRed,thick]    ([yshift=1.2mm]fo.east -| fs.west) -- ([yshift=1.2mm]fo.east);
      \draw[->,idle]              ([yshift=-1.2mm]fo.east) -- ([yshift=-1.2mm]fo.east -| fs.west);

      \node[font=\scriptsize,align=center] at (1.2,-4.3) {(a) feedforward (directed)};
    \end{scope}

    \begin{scope}[xshift=5.2cm]
      \node[io]   (rio) at (0,0)      {IO};
      \node[core] (rc1) at (0,-1.1)   {C$_1$};
      \node[core] (rc2) at (0,-2.2)   {C$_2$};
      \node[core] (rc3) at (0,-3.3)   {C$_3$};
      \node[sw]   (rs)  at (2.4,-1.65) {SW};

      \foreach \n in {rio,rc1,rc2,rc3}{
        \draw[->,thick] ([yshift=1.2mm]\n.east) -- ([yshift=1.2mm]\n.east -| rs.west);
        \draw[->,thick] ([yshift=-1.2mm]\n.east -| rs.west) -- ([yshift=-1.2mm]\n.east);
      }
      \node[font=\scriptsize,align=center] at (1.2,-4.3) {(b) recurrent (broadcast)};
    \end{scope}
  \end{tikzpicture}
  \caption{The two configurable routing modes over the \textsc{Apeiron} switch, introduced by
  the refined communication IP. Every core port is bidirectional (two arrows per link).
  (a)~Feedforward: the input core injects to C$_1$ (green leg, I$\to$SW$\to$C$_1$), which 
  routes via the switch through C$_2$ before forwarding to the output core (red legs); unused legs are greyed. 
  Delivery is directed and point-to-point, though it still traverses the switch. (b)~Recurrent: a single I/O
  core and the compute cores all exchange over the switch, and each core's output is broadcast to
  all cores, supporting arbitrary cyclic connectivity. The same cores and switch serve both
  modes; only the addressing policy changes.}
  \label{fig:topologies}
\end{figure}

The refined communication IP also makes the way cores address one another over the routing IP
configurable, following the connectivity of the mapped network (Fig.~\ref{fig:topologies}). In
\emph{feedforward} operation each core sends its output spikes only to the cores that host
postsynaptic targets, using per-core routing tables resolved at generation time; delivery is
directed (point-to-point) over the switch, and the receiving core's presynaptic address space
stays compressed to its few upstream sources, keeping the occupied synaptic memory small.
Directed delivery combined with a layer-per-core mapping also relieves the input port: a core
in a feedforward pipeline ingests only the spikes of its single upstream layer, rather than
contending for the same port between an upstream source and its own re-injected output as a
broadcasting core does, so per-core input bandwidth translates more directly into throughput.
In \emph{recurrent} operation each core broadcasts its output over the switch to all cores, the
general case for arbitrary cyclic connectivity, and the presynaptic address space expands to the
global source set. The two modes share the same cores and the same switch; only the addressing
policy and the resulting per-core memory sizing differ.

For the networks recurrent operation targets, broadcast is less wasteful than it appears. In a
sparsely-connected balanced network distributed across a modest number of cores, a neuron's
spike is likely to have at least one postsynaptic target on most cores, so directed delivery
would in any case replicate the spike widely; at the scales considered here broadcast and
directed delivery converge, while broadcast avoids per-source routing state.

\subsection{Fused single-word spike/sync protocol}
\label{sec:ref:sync}
The prototype spends a dedicated packet on the per-timestep barrier
(Section~\ref{sec:arch:system}). Moving the communication IP from a payload-carrying format to a
header-only, single-word format lets the end-of-timestep marker ride on the last spike-carrying
word, removing the dedicated synchronization transaction; the gain is largest for
throughput-bound feedforward inference, where synchronization would otherwise be a fixed
per-timestep tax.

\subsection{Neighbor-local synchronization and stochastic input}
\label{sec:ref:gals}
The current global barrier, in which every core exchanges a sync with every other, scales poorly
with core count. Restricting synchronization to nearest neighbors on the torus, a
globally-asynchronous locally-synchronous (GALS) scheme, removes the all-to-all exchange while
preserving per-timestep spike ordering. Independently, a Poisson spike generator embedded in each
worker can supply the background stochastic activity that bio-inspired networks require, without
carrying it over the fabric.

\subsection{Design-space characterization}
\label{sec:ref:char}
Once these refinements are in place, the more interesting object of study is the design space
itself, which the configuration axes make directly measurable. The same network can be
regenerated across the numeric-format, folding, banking-degree ($P$), and worker-count axes,
and each resulting instance placed and measured on the same device, turning qualitative
architectural trade-offs into a Pareto surface of accuracy against area, latency, and energy.
A full fixed-point field-width sweep maps the accuracy-versus-cost front of a single workload;
sweeping the folding and banking axes at fixed accuracy maps the latency-versus-area front;
and running the same two workload classes through the identical flow lets the feedforward and
recurrent operating points be compared on one substrate rather than across separate systems.

Energy anchors this surface. From switching activity, representative stimulus is simulated to
produce a switching-activity record (SAIF) that drives the vendor power-analysis tool to yield
average power for a configured instance; energy per inference (feedforward) or per simulated
biological second (recurrent) then follows from the measured runtime. Reported in neuromorphic
conventions, energy per synaptic operation (\si{\pico\joule}/SOP) and energy per inference
(\si{\micro\joule}/inference), these figures are read as relative comparisons across AIGOR
configurations, letting the design space be ranked by efficiency rather than compared as
absolute claims against ASIC neuromorphic processors.

\FloatBarrier
\section{Conclusion}
\label{sec:conclusion}

We have presented AIGOR, a modular, event-driven neuromorphic architecture for configurable SNN
inference. AIGOR treats the neuron model, numeric precision, and the mapping of neurons onto
workers and cores as configuration axes of one architecture rather than as design-time
commitments, and it derives a complete synthesizable instance, cores, neuron kernels, and
initialized synaptic-memory images, from a single declarative specification. A first prototype on
the Versal VPK180 runs two workloads usually handled by separate systems, a feedforward snnTorch
classifier and a recurrent NEST balanced random network, on the same cores: the classifier
reproduces its reference accuracy, and the recurrent network matches its NEST reference at the
spike level across cores spanning two FPGAs. We characterized the prototype's resource
utilization and validated the multi-node synchronization scheme in simulation up to one thousand
cores. The measured throughput localizes the bottleneck in the synaptic-delivery datapath and the
global barrier, and motivates a banked synaptic accumulator, a fused spike/sync protocol, and
neighbor-local synchronization, now in development, each expressible as a change to the same
configurable cores. The architecture's parameterized structure makes it a natural platform for
the systematic exploration of SNN hardware design trade-offs.

\section*{Acknowledgments}
P.~Perticaroli is a PhD student enrolled in the National PhD program in Artificial Intelligence,
XXXIX cycle, course on Health and Life Sciences, organized by Università Campus Bio-Medico di
Roma.


\end{document}